\documentclass[twocolumn,
	aps, prd,
	10pt, notitlepage, 
        floats, floatfix,
	amsmath, amssymb, amsfonts, eqsecnum,
	superscriptaddress,
	showpacs, showkeys,
	nofootinbib,
 	longbibliography,
]{revtex4-2}

\usepackage{graphicx}
\usepackage{dcolumn}
\usepackage{bm}
\usepackage{multirow}
\usepackage{color}
\usepackage{changepage}
\usepackage[colorlinks,urlcolor=blue,citecolor=blue,linkcolor=blue]{hyperref}
\usepackage{xcolor}
\usepackage{soul}
\usepackage{physics,amsmath}
\usepackage{subcaption}
\usepackage{hyperref}
\usepackage{array}
\usepackage{verbatim}
\usepackage{booktabs}
\usepackage{mathrsfs}
\usepackage{color}
\usepackage{enumitem}
\usepackage{xcolor}
\usepackage{tikz}
\usepackage{orcidlink}
\usetikzlibrary{decorations.pathmorphing}
\usetikzlibrary{arrows}

\newcommand{\p}{\partial}

\newcommand{\nb}{n_{\rm B}}

\newcommand{\ep}{\varepsilon}

\newcommand{\csqeq}{c_{\rm eq}^2}
\newcommand{\csqad}{c_{\rm ad}^2}

\newcommand{\csqdy}{c_{\rm dy}^2}

\begin{document}


\title{
Oscillations of Dissipative Neutron Stars: The Impact of Hyperonic Reaction Rates}

\author{Suprovo Ghosh\,\orcidlink{0000-0002-1656-9870}}
\email{S.Ghosh@soton.ac.uk}
\affiliation{
Mathematical Sciences and STAG Research Centre, University of Southampton, Southampton SO17 1BJ, United Kingdom
}

 \author{Alexander Haber\,\orcidlink{0000-0002-5511-9565}}
\email{A.Haber@soton.ac.uk}
\affiliation{
 Mathematical Sciences and STAG Research Centre, University of Southampton, Southampton SO17 1BJ, United Kingdom
}

\author{Nils Andersson \,\orcidlink{0000-0001-8550-3843}}
\email{N.A.Andersson@southampton.ac.uk}
\affiliation{
Mathematical Sciences and STAG Research Centre, University of Southampton, Southampton SO17 1BJ, United Kingdom
}

\author{Andrew Rhys Counsell \,\orcidlink{0000-0002-4532-7440}}
\email{A.R.Counsell@soton.ac.uk}
\affiliation{
Mathematical Sciences and STAG Research Centre, University of Southampton, Southampton SO17 1BJ, United Kingdom
}

\date{August $10^{\text{th}}$, 2026}

\newcommand{\rc}[1]{\textsf{\color{blue}{ #1}}}
\newcommand{\sg}[1]{\textsf{\color{red}{ #1}}}
\newcommand{\na}[1]{\textsf{\color{green}{ #1}}}

\begin{abstract}
Tidal excitations of stellar oscillation modes during binary neutron-star inspirals offer a powerful probe of the composition of dense matter at supranuclear densities. Chemical equilibration plays a crucial, but often neglected, role in stellar perturbation calculations. If the chemical equilibration timescale is comparable to the oscillation timescale, then viscous effects can damp the modes. If the reactions are fast, some modes can completely disappear since their restoring force vanishes. Typically, these calculations, however, assume either instantaneous chemical equilibrium or no equilibration (frozen composition). Motivated by this, we investigate the effects of finite reaction rates on the oscillation spectrum of neutron stars containing hyperonic matter. We calculate the dominant non-leptonic weak interaction rates and incorporate them into the relativistic perturbation equations through a complex, frequency-dependent dynamical sound speed. We show that finite-rate effects naturally manifest as bulk-viscous dissipation, modifying the properties of both the fundamental ($f$) and gravity ($g$) modes. We further examine the impact on the tidal response by matching stellar perturbations to near-zone boundary conditions, demonstrating how viscous dissipation gives rise to a tidal lag. These results provide a consistent framework connecting microscopic reaction rates and the resulting bulk viscosity to the tidal dynamics of compact binaries, and represent a step towards incorporating viscous dissipation into gravitational-wave models of binary neutron-star inspirals.
\end{abstract}

\maketitle


\section{Introduction}
\label{sec:intro}

Understanding the properties of matter at supranuclear densities inside neutron star cores remains one of the central challenges in fundamental physics. The detection of gravitational waves from the binary neutron star merger GW170817~\cite{Abbott2017,MultiGW170817} marked a turning point, enabling measurements of tidal effects during the late inspiral phase~\cite{Abbott2018}. Measurements of the tidal deformability have already provided valuable constraints on the equation of state (EOS) of neutron star matter~\cite{De_2018,Abbott2019}. The tidal deformability, like the star’s mass and radius, only depends on the equilibrium EOS of dense matter~\cite{Hinderer}. However, different microscopic compositions---which may include hyperons, deconfined quark matter, or other exotic phases---may lead to similar static macroscopic observables~\cite{Ozel:2016oaf,Oertel_2017,Baym_2018}. This is known as the ``masquerade problem"~\cite{Alford_2005}. To distinguish between different compositions, it is necessary to identify observables that are directly sensitive to the underlying degrees of freedom. \\

Dynamical tides during the binary inspiral, associated with the excitation of the stellar oscillation modes, provide an additional probe of the neutron star interior~\cite{Pnigouras:2025muo,Andersson:2019ahb,Yu_2017,Steinhoff:2016rfi,PhysRevResearch.3.033129,fmode_pratten}. The dominant contribution to the dynamical tides comes from the star's fundamental ($f$) mode, whose effect has already been included in state-of-the-art waveform models, e.g.~\cite{NRttidalv3}. At lower frequencies, gravity ($g$) modes may also be resonantly excited~\cite{Lai:1993di,Counsell:2025hcv}. These oscillations are restored by buoyancy and arise from composition gradients within the stellar interior. Although their overall contribution to the tidal phase is weaker than that of the $f-$mode~\cite{Lai:1993di,Counsell_2024}, their direct link to the composition inside neutron stars makes them high-profile targets, specifically for the next generation of gravitational-wave detectors~\cite{Gittins:2026ntx,Counsell:2025hcv,10.1093/mnras/stab1898,lzgx-qls1} (Cosmic Explorer~\cite{CE,evans2023cosmicexplorersubmissionnsf} in the USA and the Einstein Telescope~\cite{2026JCAP...03..081A} in Europe).\\

Most studies of neutron-star oscillations rely on the frozen-composition approximation~\cite{Detweiler1985}. In this limit, weak interactions that maintain chemical equilibrium are assumed to be too slow to modify the local composition of a fluid element during an oscillation cycle. The composition therefore remains fixed as the fluid oscillates, producing a difference between the equilibrium and adiabatic fluid responses that acts as a source of buoyancy. This approximation is well justified for cold nucleonic matter, where chemical equilibrium is maintained primarily by Urca reactions whose timescales are many orders of magnitude longer than the typical oscillation periods~\cite{Sawyer1989,Alford_NWA_2024}. As a result, the oscillation spectrum is accurately described by the standard perturbation framework. However, the situation changes in the presence of exotic constituents such as hyperons. In hyperonic matter, strangeness-changing non-leptonic weak interactions can operate on timescales comparable to the oscillation periods~\cite{Lindblom2002,Ghosh:2023vrx,Alford2021} even at temperatures of $10^6$–$10^9$ K, characteristic of neutron stars during the inspiral phase~\cite{Lai:1993di,Ghosh:2023vrx}. When the reaction and oscillation timescales become comparable, neither the frozen-composition nor the instantaneous-equilibrium limit provides an adequate description. Instead, the composition responds dynamically to the oscillation, leading to a phase lag between density and composition perturbations. At a macroscopic level, this lag manifests itself as bulk-viscous dissipation~\cite{Andersson_2019,Most:2022yhe,Zhao:2025pgx}. \\

Recent work~\cite{Ghosh:2023vrx} has highlighted the possibility that bulk-viscous dissipation in hyperonic matter can impact the inspiral dynamics of neutron-star binaries and produce potentially observable signatures in next-generation gravitational-wave detectors. Motivated by this prospect, several recent studies have attempted to incorporate dissipation into relativistic stellar perturbation theory through viscous corrections to the stress-energy tensor, introducing bulk- and shear-viscous transport coefficients at the level of the fluid equations~\cite{bussires2026axialoscillationsviscousneutron,Keeble_2026,Redondo_Yuste_2025,Katagiri:2026jgp}. An alternative and more microscopic approach derives bulk viscous dissipation directly from finite chemical equilibration rates~\cite{Andersson_2019,Most:2022yhe,Alford:2023gxq}. This approach has recently been explored for $g-$modes in nucleonic matter~\cite{Zhao:2025pgx}.\\

In this work, we incorporate finite chemical relaxation timescales in the general-relativistic perturbation framework through a complex, frequency-dependent dynamical sound speed. This formulation provides a unified description of composition-driven buoyancy and weak-interaction-driven bulk viscosity within a single perturbative framework. As an application, we consider neutron stars containing hyperonic matter and investigate how non-leptonic weak interaction rates modify the spectrum of fundamental ($f$) and composition ($g$) modes. We show that finite reaction rates naturally lead to bulk-viscous damping, altering the damping times primarily, and causing the eventual disappearance of the composition $g-$modes via dissipative effects. We further compute the frequency-dependent tidal response by matching the interior perturbation solution to the near-zone exterior spacetime, thereby establishing a direct connection between microscopic reaction rates, viscous dissipation, and the tidal lag that characterizes the tidal response. These results provide an important step towards incorporating dissipative tidal effects into gravitational-wave models of binary neutron-star inspirals.\\

The structure of the paper is as follows: in Section ~\ref{sec:c_dyn}, we introduce the dynamical sound speed that incorporates the finite reaction rates and its relation to the bulk viscosity.  Section \ref{sec:modestheory} outlines the stellar perturbation theory and how the relevant physics enters the equations. Section \ref{sec:results} then presents the results on how finite reaction rates from hyperons impact the $f-$ and $g-$modes, as well as the tidal response. Finally, Section \ref{sec:discussion} summarizes the work and presents ideas for future continuation of this effort.

\section{Dynamical sound speed and bulk viscosity}
\label{sec:c_dyn}
\subsection{Bulk Viscosity}
The particle content of the system we are considering is given by the nucleons (neutrons and protons), the $\Lambda$ and $\Xi^-$ hyperons, and, in the leptonic sector, electrons and muons. We are using a relativistic mean field model to consistently calculate the EOS, the relativistic dispersion relations of the baryons that enter the rate calculations, the particle composition, and finally the various sound speeds that are used in the mode calculations. Concretely, we are using the GM1'B EOS described in detail in Ref.~\cite{Gusakov:2014ota}, which satisfies astrophysical constraints on neutron star radii and maximum masses. In principle, the pressure is given as a function of the temperature $T$, and the six a priori independent particle chemical potentials $\mu_i$ (or, equivalently, the particle fractions $x_i=n_i/n_B$ with the particle densities $n_i$ and the baryon density $n_B$):
\begin{equation}
    P=P(T,x_n,x_p,x_\Lambda,x_\Xi,x_e,x_\mu) \, .
\end{equation} 
Given that we will focus on inspiral temperatures, we will neglect any influence of the temperature on the EOS, and all derivatives are taken at constant $T$. We will furthermore perform our calculations at a given baryon density $n_B=(n_n+n_p+n_\Lambda+n_\Xi)$, and enforce charge neutrality via $n_p-n_e-n_\mu-n_\Xi=0$. Our ansatz for the pressure assumes that all particle species are independent. However, the strong interaction leads to a correlation between the $\Lambda$ and $\Xi$ hyperons and the proton via the equilibration channel
\begin{equation}
    p+\Xi^-\leftrightarrow\Lambda+\Lambda\, ,
\end{equation}
which we assume to be always in equilibrium. Detailed balance then relates the chemical potentials:
\begin{equation}\label{eq:strong}
    \mu_{\Xi^-}=-\mu_p+2\mu_\Lambda \, ,
\end{equation}
which allows us to express the pressure as a function of the strangeness fraction \begin{equation}
    x_s=\frac{n_\Lambda+2n_\Xi}{n_B} \, .
\end{equation}
Overall, we have now reduced the free parameters to five, yielding
\begin{equation}
    P=P(T, n_B,x_s,x_e,x_\mu) \, ,
\end{equation}

\begin{figure}[h]
    \centering
    \includegraphics[width=\linewidth]{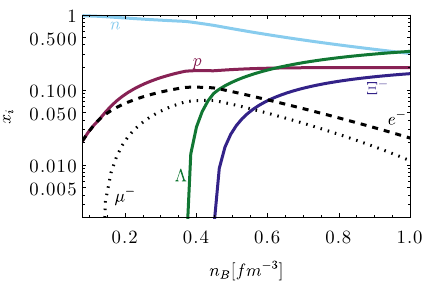}
    \caption{Particle fractions $x_i=n_i/n_B$ of the GM1'B EOS at $T=0$ in chemical equilibrium. }
    \label{fig:eos}
\end{figure}
At low temperatures, the non-leptonic decay rates that change the strangeness by one unit dominate the interactions and bulk viscosity. The possible decay channels are given by
\begin{align}\label{eq:reactions}
    n+\Lambda&\leftrightarrow p+\Xi^- \, , \nonumber \\
    n+p&\leftrightarrow p+\Lambda \, , \nonumber \\
    n+n&\leftrightarrow n+\Lambda \, , \\
    n+\Lambda&\leftrightarrow \Lambda+\Lambda \, , \nonumber \\
    n+\Xi^-&\leftrightarrow \Lambda+\Xi^- \, .\nonumber
\end{align}
 All other reactions, including the semi-leptonic Urca decays with and without hyperons \cite{Prakash:1992zng, Yakovlev:2000jp} and purely leptonic \cite{Alford:2010jf} processes, are slow at sub-MeV temperatures, so we assume them to be frozen.
Via detailed balance, we can see that all reactions apart from the first one equilibrate the difference between the neutron and Lambda chemical potential, $\delta\mu_2=\mu_n-\mu_\Lambda$. The first reaction equilibrates $\delta\mu_1=\mu_n+\mu_\Lambda-\mu_p-\mu_\Xi$. However, using Eq.~(\ref{eq:strong}), we can see that all reactions equilibrate the same chemical potential difference, $\delta\mu_1=\delta\mu_2\equiv\delta\mu$.
Instead of using the strangeness fraction, we can now alternatively use the deviation from chemical equilibrium $\delta\mu$ as a parameter, giving us $P=P(T, n_B,\delta\mu,x_e,x_\mu)$. A local density oscillation will lead to a change in pressure, for which we approximate the Lagrangian perturbation as
\begin{align}\label{eq:DP}
    \Delta P =& \left.\frac{\p P}{\p n_B}\right|_{T,\delta\mu,x_e,x_\mu}\Delta n_B+
   \left. \frac{\p P}{\p \delta\mu}\right|_{T,n_B,x_e,x_\mu}\Delta\delta\mu  \\
    &+ 
    \left.\frac{\p P}{\p x_e}\right|_{T,n_B,\delta\mu,x_\mu}\Delta x_e+ 
   \left. \frac{\p P}{\p x_\mu}\right|_{T,n_B,\delta\mu,x_e}\Delta x_\mu \, .\nonumber
\end{align}
At keV temperatures, the processes changing the lepton fractions are practically frozen, so we can neglect the third and fourth terms in this expansion. We will, from now on, assume that all derivatives are taken at constant $T,\,x_e,$ and $x_\mu$ which means that we have
\begin{equation}
    \Delta P =  \left.\frac{\p P}{\p n_B}\right|_{\delta\mu}\Delta n_B+
    \left.\frac{\p P}{\p \delta\mu}\right|_{n_B}\Delta\delta\mu \, .
\end{equation}
After defining the out-of-equilibrium strangeness fraction 
\begin{equation}
    x_{\delta s}=x_s-x_{\text{eq}}\, ,
\end{equation}
with the equilibrium fraction $x_{\text{eq}}$ being reached when $\delta\mu=0$, we can expand a perturbation in $\delta\mu$ as
\begin{equation}
    \Delta\delta\mu=\left.\frac{\p\delta\mu}{\p n_B}\right|_{\delta x_s}\Delta n_B + \left.\frac{\p\delta\mu}{\p \delta x_s}\right|_{n_B}\Delta x_{\delta x_s} \, ,
\end{equation}
where we have made the same approximations and notational choices as in Eq.~(\ref{eq:DP}). Equilibrium is restored via a change in $x_{\delta s}$ driven by the linearized net-rate $\lambda$, given by
\begin{equation}\label{eq:lambda}
    \lambda=\left.\frac{\p\left(\Gamma_\rightarrow - \Gamma_\leftarrow\right)}{\p\delta\mu}\right|_{\delta\mu=0} \, ,
\end{equation}
where $\Gamma_\rightarrow$ and  $\Gamma_\leftarrow$ are the forward/backward rates of the reactions of Eq.~(\ref{eq:reactions}), and $\lambda$ is given by the sum of the individual values, $\lambda=\sum_{i=1}^5\lambda_i$ where $i$ runs over the five reaction channels from Eq.~(\ref{eq:reactions}).
The change in the strangeness fraction can then be written as
\begin{equation}
    \frac{\p x_{\delta s}}{\p t}=\frac{\lambda}{n_B}\delta\mu \, .
\end{equation}
We can use this result to calculate the temporal change in $\Delta\delta\mu$ as
\begin{align}
        \frac{\p\Delta\delta\mu}{\p t} &= \left.\frac{\p\delta\mu}{\p n_B}\right|_{\delta x_s}\frac{\p\Delta n_B}{\p t} + \left.\frac{\p\delta\mu}{\p \delta x_s}\right|_{n_B}\frac{\p\Delta x_{\delta x_s}}{\p t} \nonumber \\ 
        &= \left. \frac{\p\delta\mu}{\p n_B}\right|_{\delta x_s}\frac{\p\Delta n_B}{\p t} -\gamma\Delta\delta\mu \, ,
\end{align}
where we have defined the inverse equilibration time 
\begin{equation}\label{eq:gamma}
    \gamma \equiv -\frac{\lambda}{n_B}\left.\frac{\p\delta\mu}{\p x_{\delta s}}\right|_{T,n_B,x_e,x_\mu} \, .
\end{equation}
Assuming that all perturbations $\Delta A$ are harmonic, \mbox{$\Delta A = \Delta A e^{i\omega t}$}, we obtain
\begin{equation}
    \Delta\delta\mu=\frac{\frac{\p\delta\mu}{\p n_B}\Delta n_B}{1+\frac{\gamma}{i\omega}} \, .
\end{equation}
We are now able to calculate $\Delta P$ from Eq.~(\ref{eq:DP}) as
\begin{align}
    \Delta P &= \left( \frac{\p P}{\p n_B} + \frac{\p P}{\p \delta \mu}\frac{\frac{\p\delta\mu}{\p n_B}}{1+\frac{\gamma}{i\omega}}\right)\Delta n_B \nonumber \\
    &= \frac{P\Gamma}{n_B}\Delta n_B \, ,
\end{align}
where the complex-valued effective adiabatic index for a damped oscillation of frequency $\omega$ is given by 
\begin{eqnarray} \label{eq:Gamma_dy}
\Gamma &=&  \dfrac{\nb}{P} \left[\left.\dfrac{\partial P}{\partial \nb}\right|_{{\delta\mu}}+\dfrac{\left.\dfrac{\partial P}{\partial {\delta\mu}}\right|_{\nb} \left.\dfrac{\partial {\delta\mu}}{\partial \nb} \right|_{\delta x} }{1+ \dfrac{\gamma}{i\omega}}\right]. 
\end{eqnarray}
In analogy with the nuclear derivation in Ref.~\cite{Zhao:2025pgx}, we use this expression to define the dynamic sound speed
\begin{equation}
    \csqdy=\frac{P}{P+\varepsilon}\Gamma \, ,
\end{equation}
with the energy density $\varepsilon$. This complex quantity allows us to capture the effects of the finite interaction rates in the mode equations.
The real and imaginary parts of the dynamical sound speed squared are then given by
\begin{eqnarray}
{\rm Re}\left[\csqdy\right] &=& \csqeq+\left(\csqad-\csqeq\right)\dfrac{\omega^2}{\omega^2+\gamma^2} \, ,\label{eq:csq_dy_real}\\
{\rm Im}\left[\csqdy\right] &=& \left(\csqad-\csqeq\right)\dfrac{\omega\gamma}{\omega^2+\gamma^2}\, . \label{eq:csq_dy_imag}
\end{eqnarray}
The equilibrium sound speed can be read off from Eq.~(\ref{eq:Gamma_dy}) and is given by
\begin{equation}
    \csqeq=\frac{n_B}{P}\left.\frac{\partial P}{\partial n_B}\right|_{\delta\mu,T,x_\mu,x_e}=\frac{P}{P+\varepsilon}\Gamma_\text{eq} \, \label{eq:cseq} .
\end{equation}
Note that this is not the usual equilibrium speed of sound (${\tilde{c}_{\rm eq}^2} = dP/d\varepsilon$) which is traditionally calculated in chemical equilibrium w.r.t.~the electron and muon fractions. Since Urca reactions, which could change these particle fractions, are always frozen at inspiral temperatures on the timescales set by the $f-$ and $g-$mode frequencies, we calculate the speed of sound only in strangeness equilibrium. Hence, we will use the expression in Eq.~(\eqref{eq:cseq}) as our equilibrium sound speed in this calculation.
The adiabatic speed of sound is given by
\begin{equation}
   \csqad =\frac{n_B}{P}\left.\frac{\partial P}{\partial n_B}\right|_{\delta x_s,T,x_\mu,x_e}=\frac{P}{P+\varepsilon}\Gamma_\text{ad} \, . \label{eq:csad}
\end{equation}

The imaginary part, $ {\rm Im}\left[\csqdy\right]$, is directly related to the bulk viscosity coefficient $\zeta$, which is defined via the energy dissipation rate  $\mathop{{\rm d}\ep}/\mathop{{\rm d}t}=-\zeta(\nabla\cdot\mathbf{v})^2$, since the bulk viscosity coefficient is also equal to the EOS-related factor times the resonance expression $\gamma/\left(\gamma^2+\omega^2\right)$~\citep{Harris:2024evy}. Using thermodynamic identities and the definitions of the adiabatic index from Eq.~(\ref{eq:Gamma_dy}), the bulk viscosity can be written as

\begin{equation}
\zeta =\dfrac{\ep+P}{\omega}{\rm Im}\left[\csqdy\right] \, .
\label{eq:bv}
\end{equation}
\subsection{Rate Calculation}
Several channels contribute to the rate $\gamma$ in the dynamical speed of sound. In the past, the matrix element for the rate has often been calculated in terms of a contact interaction derived from W-boson exchange. In Ref.~\cite{vanDalen:2003uy}, it has been shown that a combined strong-weak vertex leads to significantly faster strangeness equilibration channels and allows for additional decay channels to open up. Since equilibration and bulk viscosity will be dominated by the fastest equilibration channels, we will neglect the contact interaction channels and instead use the combined strong-weak interaction channels. For the corresponding Feynman diagrams and quark-flow diagrams, see Ref.~\cite{Alford2021} and \cite{Ofengeim:2019fjy}. Given the low temperatures, we are relying on the Fermi-surface approximation, where all degenerate particles are fixed to their Fermi surface. In this approximation, the slope of the difference of the different rate channels $\lambda_i$ can be described by an analytical temperature dependence of $\lambda_i\propto T^2$. We derive the rate in App.~\ref{App:rate}, and only quote the final result here:
\begin{equation}
    \label{eq:rate_final}
    \lambda_i= \frac{T^2}{6144\pi^6}\mathcal{A}_i  \,,
\end{equation}
with the angular integral $\mathcal{A}_i$  derived in App.~\ref{App:rate}.
\subsection{Dynamical Speed of Sound}

\begin{figure}[h]
    \begin{subfigure}{0.5\textwidth}
        \includegraphics[width=\textwidth]{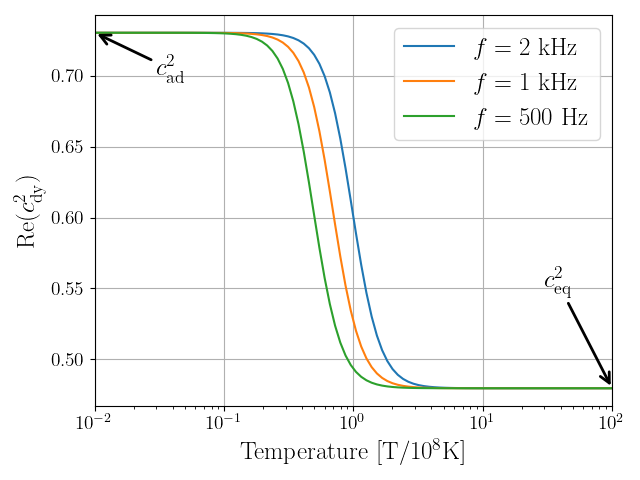}
    \end{subfigure}
    \begin{subfigure}{0.5\textwidth}
        \includegraphics[width=\textwidth]{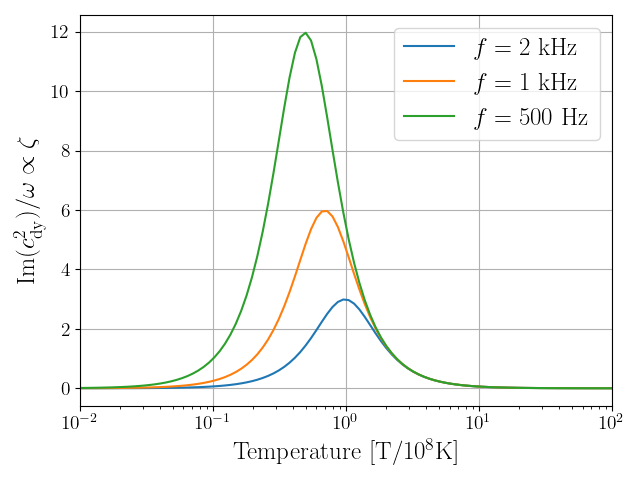}
    \end{subfigure}
    \caption{Real (upper panel) and imaginary part (lower panel) of the dynamical sound speed ($\csqdy$) as a function of temperature at $n_B = 0.6\ \mathrm{fm}^{-3}$ at three frequencies $f$. }
    \label{fig:cdyn}
\end{figure}
The real and imaginary parts of the dynamical sound speed are plotted in Fig~\ref{fig:cdyn}, as a function of temperature for a fixed density, but different input frequencies. We can see that the real part of $\csqdy$ ranges from $\csqad$ to $\csqeq$ with increasing temperature, and hence the reaction rates, according to Eq.~(\eqref{eq:rate_final}). Once the dynamical speed of sound approaches the equilibrium speed of sound, composition-driven modes (i.e. $g-$modes) vanish due to the missing restoring force. Physically, a displaced fluid element will immediately chemically equilibrate to its environment and thus not experience a restoring displacement force. Before this regime is reached, viscous effects can dampen the modes as well. This can be seen from the imaginary part, related to the bulk viscosity according to Eq.~(\ref{eq:bv}), which shows a frequency-dependent peak at a finite temperature. This means that the bulk viscous impact on different modes with different frequencies is expected to be different.

\section{Quasi-normal Modes}
\label{sec:modestheory}
For the calculation of the neutron star's quasi-normal mode spectrum, we  follow the perturbation formulation developed by Detweiler and Lindblom~\cite{Detweiler1985} and the augmentation introduced by Krüger et al.~\cite{Kruger_2015} to mitigate numerical noise in the low-frequency spectrum. We consider linear oscillations of a nonrotating, spherically symmetric relativistic star.  The metric tensor that describes the equilibrium background is given by
\begin{equation}
ds^2 = -e^{\nu(r)} dt^2 + e^{\lambda(r)} dr^2 + r^2(d\theta^2+\sin^2\theta\,d\phi^2)\, ,
\end{equation}
together with a perfect fluid stress-energy tensor 
\begin{equation}
   T^{\mu\nu}=(\varepsilon+p)u^\mu u^\nu + p g^{\mu\nu}\,.  
\end{equation}
\noindent
We introduce linear Eulerian perturbations ($h_{\mu\nu}$) to this metric,
\begin{equation}
g_{\mu\nu} \rightarrow g_{\mu\nu} + h_{\mu\nu},
\end{equation}
along with the Lagrangian fluid displacement vector $\xi^\mu$, perturbations of the fluid 4-velocity($u^{\mu}$), and thermodynamical quantities $P$ and $\varepsilon$. \\
\noindent
Due to the underlying spherical symmetry, all perturbations can be decomposed into tensor spherical harmonics with angular indices $(\ell,m)$. We adopt the
Regge-Wheeler gauge, where high angular derivatives are chosen to vanish. In the polar-parity sector (quantities transform as $(-1)^{\ell}$), the metric
perturbations take the form
\begin{equation}
\begin{split}
    & h_{\mu\nu}^{\text{polar}} =- Y_{\ell m}(\theta,\phi)\, e^{i\omega t}\times\\
    &\begin{pmatrix}
e^{\nu} r^\ell H_0(r) & i\omega r^{\ell+1} {H}_1(r) & 0 & 0 \\
i\omega r^{\ell+1} {H}_1(r) & e^{\lambda} r^\ell {H}_2(r) & 0 & 0 \\
0 & 0 & r^{\ell+2} {K}(r) & 0 \\
0 & 0 & 0 & r^{\ell+2}\sin^2\theta \,{K}(r)
\end{pmatrix}
\end{split}
\label{eq:polar-r-l}
\end{equation}
\noindent
Meanwhile, using the gauge freedom with the Lagrangian formulation, we set $u^{\mu}\xi_{\mu} = 0$, and expand the non-trivial components as
\begin{equation}
\begin{split}
 \xi^r(r) &= e^{-\lambda/2}r^{\ell - 1}W(r) Y_{\ell m} e^{i\omega t}\,, \\
\xi^\theta& = -r^{\ell - 2}V(r)\partial_\theta Y_{\ell m} e^{i\omega t}\,,\\
\xi^\phi &= -\frac{r^{\ell - 2}}{\sin^2{\theta}}V(r)\partial_\phi Y_{\ell m} e^{i\omega t}\,.   
\end{split}
\end{equation}
\noindent
In this formalism, the perturbation is described by the spacetime variables $H_0(r), H_1(r), H_2(r), K(r)$, and the fluid variables $W(r), V(r)$. These five perturbation functions are, however, not all independent. Detweiler and Lindblom~\cite{Detweiler1985} reduced the pulsation equations (not explicitly shown here) for the interior of the star to a system of four first-order differential equations for the functions $H_1(r), K(r), W(r), X(r)$ (See Eq. 8--11 in Ref.~\cite{Detweiler1985}), where $X(r)$ is an auxiliary variable proportional to the Lagrangian pressure perturbation, defined as
\begin{align}
    X(r) &= \omega^3(p+\rho)e^{-\nu/2}V - r^{-1}p'e^{(\nu-\lambda)/2}W \nonumber\\
    &+ \frac{1}{2}(p+\rho) e^{\nu/2}H_0\, . 
\end{align}
However,  at the low frequencies relevant for $g-$modes, this perturbation formulation meets with numerical difficulties~\cite{Kruger_2015}. To alleviate this, we adopt the strategy described in Appendix B of Kruger et al.~\cite{Kruger_2015}, which essentially uses $V$ instead of $X$ as an independent variable.\\

\noindent
The four independent perturbation equations are solved subject to two boundary conditions. 
\begin{enumerate}
    \item  At the center of the star, $r = 0$, the solutions must be regular. They are determined by using the Taylor series expansion method described in Section II of Ref. \cite{Detweiler1985}. 
    \item The perturbed pressure should vanish at the stellar surface, which is equivalent to the condition $X(R) = 0$.
\end{enumerate} 
These boundary conditions are sufficient to determine ($H_1, K, W, X$) in the stellar interior for any given frequency, up to an arbitrary amplitude.\\
\noindent
To close the fluid perturbation problem inside the star, we need to provide a matter equation of state and a relation between perturbed (Lagrangian) pressure($\Delta P$) and energy density ($\Delta \varepsilon$). This is where the dynamical sound speed enters the perturbation equations. As shown in Sec.~\ref{sec:c_dyn}, in the most general scenarios where we have finite reaction rates, we have
\begin{equation}
\label{eq:soundspeed}
    \Delta P = \csqdy\Delta \varepsilon\,.
\end{equation}
We can quickly check the appropriate limits (as also shown in Fig.~\ref{fig:cdyn}):
\begin{itemize}
    \item when rates are slow compared to perturbation timescales (i.e.$\gamma \ll \omega$), $\csqdy \rightarrow \csqad$. This is the \textit{frozen composition} limit, where the difference $\csqad - \csqeq$ gives rise to $g-$modes.
    \item when rates are fast compared to perturbation timescales (i.e. $\gamma \gg \omega$), $\csqdy \rightarrow \csqeq$. This is the \textit{fast reaction} limit, where the system always remains in equilibrium. In this case, there will be no $g-$modes.
\end{itemize}

\noindent
Outside the stellar surface, where the fluid perturbations vanish, we only have perturbations of the spacetime. The metric perturbation functions $H_1$ and $K$ can be combined into a single variable obeying a single differential equation, known as the Zerilli equation~\cite{Zerilli}. The Zerilli equation is integrated from the stellar surface to infinity ($r \rightarrow \infty$), with the initial boundary fixed using the stellar interior perturbation solutions at the surface of the star. At infinity, the solutions correspond to both ingoing and outgoing solutions. A free oscillation mode corresponds to the solution with no ingoing waves. This defines an eigenvalue problem for the quasi-normal modes of the star.

\section{Results}
\label{sec:results}

We want to explore the impact of the dynamical sound speed, which incorporates the finite reaction rates that maintain the background chemical equilibrium, on the quasi-normal mode spectrum. As described in Sec. II, this will also quantify the weak-interaction-driven bulk viscous impact on damping the modes.  Since we are interested in observing the mode oscillations through gravitational wave observations, we restrict ourselves to the study of $g$ and $f-$modes, whose frequencies lie in the sensitivity band ($\sim 10- 2000$ Hz) of current and upcoming gravitational wave observatories.
\subsection{$f-$mode}
The fundamental $f-$mode is expected to be damped by the bulk viscous effects of the finite reaction rates, but is otherwise independent of the reaction rates. First, we consider a heavy $2M_{\odot}$ star, which will have a large hyperon core. Given that the $f-$mode is insensitive to the composition gradient, we observe no change to the real part of the frequency with the introduction of finite reaction rates. Bulk viscosity leads to shorter mode damping times, $\tau = 1/\mathrm{Im}(\omega_f)$. These strongly depend on temperature, and are tabulated in Table~\ref{tab:fmode}. 

\begin{table}[ht]
\centering
\begin{tabular}{|c|c|c|}
\hline
\textbf{Temperature} &
\textrm{\hspace*{4mm}$\tau$\hspace*{4mm} } &
\textrm{\hspace*{3mm}$\tau_\mathrm{visc}$\hspace*{3mm}} \\
($\times 10^8$ K) &  (s) & (s) \\
\hline

$0.01$ &  $0.138$ & $270$ \\
 \hline
$0.2$ & $0.115$ & $0.71$ \\
 \hline

$0.43$ &  $0.098$ & $0.33$ \\
 \hline

$1$ &  $0.123$ & $1.13$ \\
 \hline

$10$ & $0.138$ & $158$ \\
 \hline 
\end{tabular}
\caption{Damping time of $f-$mode for a 2$M_{\odot}$ star as a function of increasing temperature. Both the total damping time ($\tau$) and viscous damping time ($\tau_\mathrm{visc}$) are given. The corresponding mode frequency does not change with the reaction rates and thus the temperature, and is found to be $2154$ Hz.}
\label{tab:fmode}
\end{table}

Although the real part of the $f-$mode frequency does not change, the damping time reaches a minimum at a temperature of $\sim 4\times 10^7$ K. From Fig.~\ref{fig:cdyn},  we see that the imaginary part of the dynamical sound speed also reaches its peak close to the same temperature of $10^8$ K, for the frequency of $2$ kHz. As expected, the maximum damping occurs when the bulk viscosity is also maximum. Note that this damping time includes contributions from both gravitational wave emission from the modes, as well as viscous dissipation. Away from the bulk viscous peak, the damping time is dominated by gravitational wave emission. As we are still in the perturbative regime, different damping mechanisms are added as in a parallel resistor, 
\begin{equation}
    {1 \over \tau} = {1 \over \tau_\mathrm{GR}} + {1 \over \tau_\mathrm{visc}}\,.
    \label{eq:parallel}
\end{equation}
The damping due to gravitational wave emission does not change with temperature. The viscous damping $\tau_\mathrm{visc}$ can be calculated from Eq.~(\ref{eq:parallel}) and is reported in Table~\ref{tab:fmode}. In this setup, the viscous damping can reach the same order as the damping from gravitational wave emission, making the total damping rate faster by $30$\%.   \\

\begin{figure}[h]
    \centering
    \includegraphics[width=1\linewidth]{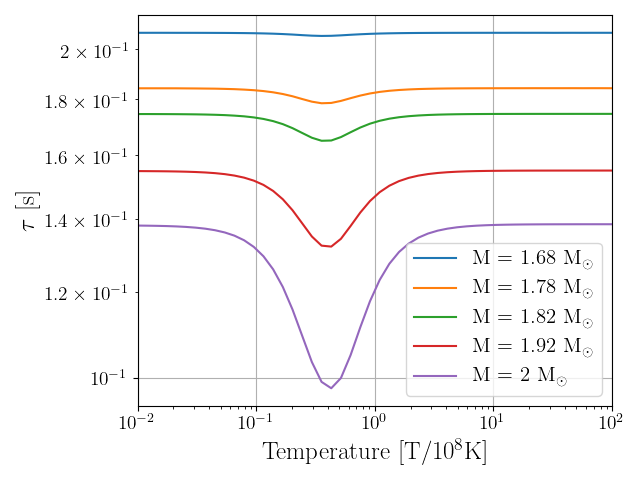}
    \caption{Damping time of the $f-$mode as a function of increasing temperature for different stellar masses.}
    \label{fig:fmode}
\end{figure}

We further investigate the influence of the stellar mass (or equivalently the compactness) of the neutron star. In Fig~\ref{fig:fmode}, we plot the damping time of the $f-$mode for increasing masses. For relatively lower masses, the central density is not high enough to support a large hyperon core. Hence, the bulk viscous contribution to the damping time is negligible. With increasing masses, we can clearly see how viscous dissipation impacts the total damping of the $f-$mode. Note that the gravitational radiation-driven damping also increases as you increase the compactness, following the universal relation reported in the literature~\cite{Andersson1997,Pradhan:2022vdf}. The fastest damping time occurs at temperatures close to the bulk viscous peak, which is weakly dependent on the EOS and hence density, and strongly depends on temperature. The observed small difference between the peak location in Fig.~\ref{fig:cdyn} and Fig.~\ref{fig:fmode} is due to the fact that the f-mode damping is an integrated quantity to which also the purely nuclear parts of the star contribute, while the bulk viscous peak is shown at a fixed single baryon density.

\subsection{$g-$mode}

\begin{figure}[h]
    \centering
    \includegraphics[width=1\linewidth]{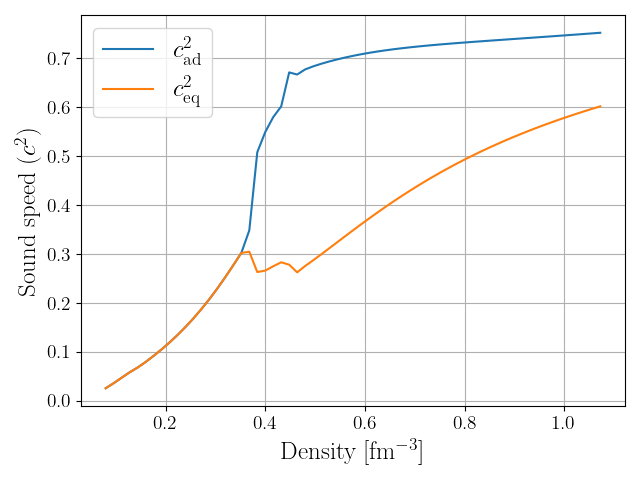}
    \caption{Adiabatic and strangeness-equilibrium sound speed from Eq.~(\ref{eq:csad}) and Eq.~(\ref{eq:cseq})  as a function of density for the GM1'B equation of state. The sound speed difference increases when hyperons start to appear at $n_B = 0.35$fm$^{-3}$.} 
    \label{fig:speed_diff}
\end{figure}
 To explore the impact of hyperons on the $g-$modes, we need to contrast the adiabatic index for the perturbations given by the dynamical sound speed ($\csqdy$) with that of the background ($\csqeq$), which is computed in strangeness equilibrium, but at frozen lepton fractions as mentioned in Sec.~\ref{sec:c_dyn}. Note that in this work, we are not considering other particle gradients, like protons \cite{1983ApJ...268..837M, 1992_Reisenegger}, which typically results in lower-frequency $g-$modes~\cite{Counsell:2023pqp}. These modes will not be influenced by the practically frozen Urca reactions at these temperatures, but rather have a constant frequency depending on the nuclear EOS. Only at temperatures in the MeV range will these modes be altered by the increasing Urca reaction rates \cite{Zhao:2025pgx}. If we consider the sound speed diagram for hyperons in Fig.~\ref{fig:cdyn}, we see that at lower temperatures ($T/10^8K \ll 0.1$), the dynamical sound speed approaches $c_\mathrm{ad}^2$. This is the frozen composition limit, where we expect to see the $g-$mode spectrum. As we increase the temperature, $\csqdy$ approaches the equilibrium limit at $T/10^8K \gg 10$, where we should not see any $g-$modes. In this transition region, we expect the biggest influence of bulk viscosity, and thus want to explore how the finite reaction rates impact the $g-$mode spectrum.\\

 In Table~\ref{tab:gmode}, we list the frequency and damping time of the first three $g-$modes as a function of increasing temperatures. When the temperature is very low ($T \ll 10^8$K), the rate is very low compared to the $g-$mode frequencies. 
 First, we notice that the $g_1-$mode frequency is actually significantly higher ($\sim 850$ Hz) than in $npe$ matter with $g_1-$mode frequencies ($\sim 100-500$ Hz~\cite{Zhao:2025pgx,Counsell_2024}). This feature has been reported in the literature~\cite{Tran:2022dva}. The same is true for higher-order $g-$modes. This can be understood from the fact that the $g-$mode frequencies are proportional to the Brunt--V\"ais\"al\"a frequency, which depends on the sound speed difference $c^2_\mathrm{ad} - c^2_\mathrm{eq}$~\cite{1992_Reisenegger,Lai:1993di}. As shown in Fig.~\ref{fig:speed_diff}, when hyperons start to appear at higher densities, they soften the equation of state, reducing  $c_\mathrm{eq}^2$ in that region. At the same time, $c_\mathrm{ad}^2$ also increases with the appearance of hyperons, causing the speed difference to jump. Hence, the $g-$mode frequencies are higher in the presence of hyperons. In principle, this could be a possible smoking gun signature for the existence of hyperons inside neutron star cores~\cite{Tran:2022dva}. \\
 
\begin{figure}[h]
    \centering
    \includegraphics[width=1\linewidth]{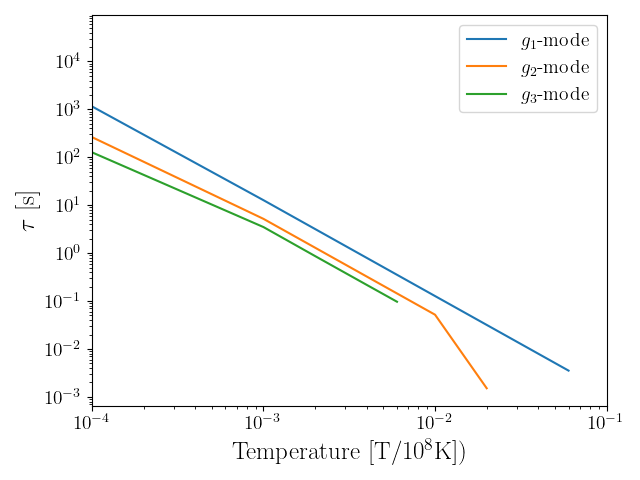}
    \caption{Damping times of the three leading-order $g-$modes as a function of increasing temperature for a $1.92M_{\odot}$ star. }
    \label{fig:gmode}
\end{figure}
\noindent
When the rates are effectively zero, at very low temperatures, the $ g$-modes will be damped only by gravitational-wave emission. The gravitational damping of $ g$-modes is known to be extremely weak, as we confirm in Table~\ref{tab:gmode}. The damping times are of the order of a few hours. But when the rates of the reactions increase due to an increasing temperature, the $g-$modes are heavily damped by finite bulk viscosity. The damping times become several orders of magnitude lower, as shown in Fig.~\ref{fig:gmode}, and can reach the order of $~\sim 1$ ms around T $\sim 10^6 - 10^7$K. We observe that the hyperon $g-$mode spectrum disappears around  $T \sim 6 \times 10^6$ K, when the longest-lasting $g_1-$mode vanishes. In principle, the modes can vanish due to overdamping by bulk viscosity, or the faster rates themselves. In this limit of instantaneous relaxation at very high temperatures, the restoring buoyant force ($\propto c^2_\mathrm{ad} - c^2_\mathrm{eq}$) vanishes (as shown in Fig.~\ref{fig:cdyn}), and thus the $g-$modes are expected to vanish. But here we see, from Table~\ref{tab:gmode} as well as Fig.~\ref{fig:gmode}, that the $g-$modes vanish at a much lower temperature (the $g_1-$mode vanishes at $T \approx 6 \times 10^6$K). This is purely due to the viscous damping, since $c_\mathrm{dyn}^2 \rightarrow c_\mathrm{ad}^2$ at this temperature. This finding is in contrast to the recent finding for nuclear matter $g-$modes in Ref.~\cite{Zhao:2025pgx} at $T \geq 1$ MeV, where $g-$modes are damped by the bulk viscosity, but survive the bulk viscous peak until the fast equilibration finally kills off the modes. For hyperons, the strength of bulk viscosity is much higher than that of Urca reaction-driven bulk viscosity, simply because the sound speed difference $ c^2_\mathrm{ad} - c^2_\mathrm{eq}$ is much higher in the case of hyperons. Hence, for hyperons, the $g-$modes are completely damped out at a lower temperature when bulk viscosity strength is still at $~\sim 10\%$ of its peak value. This is also the reason why the real part of the frequency does not change in the temperature regions where we find the $g-$modes. We see a similar trend for the higher-order $g-$modes as well, only they are damped away much faster at lower temperatures because the damping time has an inverse scaling with the mode frequency. For higher (lower) masses of the star, both the mode frequency and damping can increase (decrease) considering the larger (smaller) hyperon core inside the star.\\

In principle, the modes could reappear at higher temperatures once we are past the resonant bulk-viscous peak where the modes are damped. However, in this regime, the reaction rates are so fast that no modes are supported. As mentioned in Sec.~\ref{sec:c_dyn}, we neglect the proton fraction gradient. At the temperatures considered here, the gradient of the proton fraction would only lead to a small correction of the g-mode frequency. However, beyond the hyperon bulk viscous peak when hyperon reactions are too fast, the proton gradient would support a g-mode at lower frequency, and the modes would not disappear completely. Note that once the temperature rises to a few MeV, the Urca rates, which are negligible at inspiral temperatures, need to be taken into account.
\begin{table*}[ht]
\centering
\begin{tabular}{|c|c|c|c|c|c|c|}
\hline
\multirow{2}{*}{Temperature ($\times 10^8$K)} 
& \multicolumn{2}{|c|}{$g_1-$mode} 
& \multicolumn{2}{|c|}{$g_2-$mode} 
& \multicolumn{2}{|c|}{$g_3-$mode} \\
\cline{2-7}
 & Frequency (Hz) & \hspace*{3mm}$\tau$(s) \hspace*{3mm}& Frequency (Hz) & $\hspace*{3mm}\tau$(s) \hspace*{3mm}& Frequency (Hz) & \hspace*{3mm}$\tau$(s) \hspace*{3mm} \\
\hline
$\sim 0$ & $853$ & $47630$ & $485$  & $39653$  & $346$ & $38945$ \\
\hline
$0.001$ &  $853$ & $12.7$  & $485$ & $5.19$ & $346$ & $3.49$ \\
\hline
$0.01$ & $853$ & $0.127$ & $485$ & $0.052$ & - & - \\
\hline
$0.03$ & $853$ & $0.013$  & - & - & - & - \\
\hline
$0.05$ & $853$ & $0.005$ & - & - & - & - \\
\hline
$0.1$ & - & - & - & - & - & - \\
\hline
\end{tabular}
\caption{Frequency and damping time of first $3$ $g-$modes for a $1.92$ $M_{\odot}$ star as a function of increasing temperature. A dash indicates that there are no $g-$modes found at these temperatures.}
\label{tab:gmode}
\end{table*}

\subsection{Impact on the tidal response}
We also want to explore the impact of the finite reaction rates and bulk viscous dissipation on the dynamical tidal response of neutron stars, which impacts the gravitational waveforms used for analyzing the inspiral signals. As we have seen in the earlier sections, for hyperons, the reaction rates impact the mode properties in a temperature region of $10^6 -10^9$K, which is the expected range of temperatures of the individual neutron stars that are yet to merge~\cite{Lai:1993di,Ghosh:2023vrx}.\\

In the low-frequency, adiabatic (slowly changing tidal field) regime, the tidal response of a neutron star is encoded in the static Love number -- tidal deformability, defined as the ratio between the induced quadrupole moment and the external tidal field~\cite{Hinderer,Binnington_2009}. At finite frequencies, however, the response is no longer instantaneous. Instead, it becomes governed by the excitation of the star's internal degrees of freedom, giving rise to a frequency-dependent response function~\cite{Andersson:2019ahb,Andersson_2026}. These dynamical tidal effects are captured via the frequency-dependent effective Love number. Here we adapt a strategy developed in Ref.~\cite{Andersson_2026} that matches the internal perturbation at the weak-field ($M/r\ll 1$) near zone ($\omega r \ll 1$) limits. Remember, for the free oscillation modes, we applied the boundary condition at infinity($r \rightarrow \infty$) that there will be no incoming waves.\\

Since the finite reaction rates impact mostly the imaginary part of the mode frequency, we consider the ansatz of $H_0$ outside the star as~\cite{Lindblom_matching}
\begin{multline}
    H_0(r,\omega) = A\bigg\{[1+\alpha (r\omega)^2 ] Q_{\ell}^2(r / M - 1) \\-iK_{\ell} \left[ r \omega \frac{d}{d (r \omega)} + 1 + \frac{1}{2} \ell (\ell + 1) - (r \omega)^2 \right]  j_{\ell}(r \omega) \bigg\} \\
    + B[1+\beta (r\omega)^2 ] P_{\ell}^2(r / M - 1),
    \label{H0final}
\end{multline}
with
\begin{equation}
    K_{\ell}=\frac{2(\ell+2)!(M \omega)^{\ell+1}}{\ell(\ell-1)(2\ell-1)!!(2\ell+1)!!},
\end{equation}
where $P_{\ell}$ and $Q_{\ell}$ are the associated Legendre polynomials of the first and second kind (of degree $l$), respectively, and $j_{\ell}$ are the spherical Bessel functions. To the leading order, we require~\cite{Andersson_2026}
\begin{equation}
    \alpha= -{1\over
    2} + \mathcal O\left( {M\over r}\right) \,,
\end{equation}
and
\begin{equation}
    \beta = -{11\over 42} - {107\over 63} {M\over r} +\mathcal O\left[(M/r)^2 \right].
\end{equation}    
For a generic frequency, matching $H_0$ in equation \eqref{H0final} with the interior solutions (as described in Sec.~\ref{sec:modestheory})  at the surface $r = R$ will determine the coefficients $A$ and $B$. The effective Love number obtained, as a function of frequency, will be given by 
\begin{equation}
    k^\mathrm{eff}_2(\omega) = {4\over 15} \left({M\over R}\right)^5 \frac{A}{B}\,.
    \label{kratio}
\end{equation}
    
\begin{figure}[h]
    \centering
    \includegraphics[width=1\linewidth]{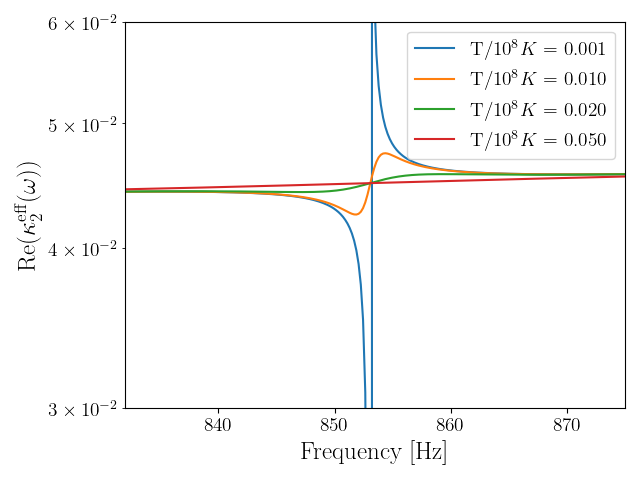}
    \caption{Effective tidal response around the $g_1-$mode frequency for $M=1.92M_{\odot}$ for increasing temperatures and thus reaction rates. Note the scale difference of the $y-$axis compared to Fig.~\ref{fig:response_f}. }
    \label{fig:response_g}
\end{figure}
\begin{figure}[h]
    \centering
    \includegraphics[width=1\linewidth]{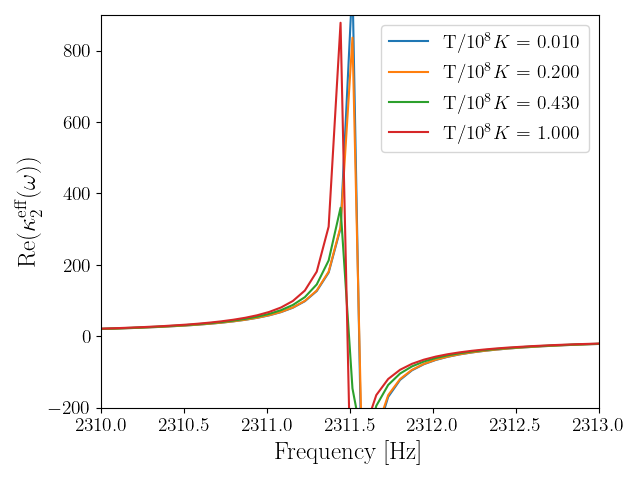}
    \caption{Effective tidal response around $f-$mode frequency for $2M_{\odot}$ with increasing reaction rates. }
    \label{fig:response_f}
\end{figure}

As shown previously~\cite{Andersson_2026}, this approach correctly converges to the static limit of the Love number and reproduces the mode resonances quite well. Here we explore how the effective Love number changes when we add the finite reaction rates. In Fig.~\ref{fig:response_g} and Fig.~\ref{fig:response_f}, we look at $k_2^\mathrm{eff}$ close to the $g_1-$mode and $f-$mode, respectively. As you increase the temperature, the $ g-$mode damping by bulk viscosity is also visible in the tidal response before it gets completely damped out close to $T \sim 6\times 10^6$ K. For the $f-$mode, we see that the response strength is diminished by viscous damping and is minimal when the viscous damping reaches a maximum at $T \sim 4\times 10^7$ K. In conclusion, bulk viscous damping tends to weaken the resonant mode excitation. The damping may also impact the accumulated tidal lag, even below resonance, as discussed below.

For a better understanding of the tidal response close to the resonances, we consider the ansatz for the tidal response, introduced in Ref.~\cite{Andersson_2026} 
\begin{equation}
        k^\mathrm{eff}_{l} = - \frac{2 \pi }{2 l + 1 }
            \sum_{n} {\mathcal A_n  \over 
            (\tilde \omega + \tilde \sigma_n-i \tilde \gamma_n)(\tilde \omega- \tilde \sigma_n-i \tilde \gamma_n)} \, ,
        \label{ksum}    
\end{equation}
where $\mathcal A_n $ replace the Newtonian mode overlap integrals, $\sigma_n$ and $\gamma_n$ are the real and imaginary parts of the mode frequencies, respectively. For any realistic neutron star model, the tidal response is dominated by the $f-$mode, i.e. $\mathcal{A}_f \gg \mathcal{A}_g$~\cite{Andersson:2019ahb,Andersson_2026}. Assuming that we are also away from the $f-$mode resonance (a reasonable assumption for inspiralling neutron stars as the $ f-$mode frequency is $\sim 2$ kHz), we can quantify  the impact of viscous damping on the tidal response by the single parameter \textit{tidal lag} ($\tau_d$), defined as
\begin{equation}
    Q_2 =  \frac{2}{3}R^{5}k_2(\chi - \tau_d\dot{\chi})\, ,
\end{equation}
where $Q_2$ is the quadrupole moment, $k_2$ is the $l = 2$ Love number, and $\chi$ is the induced tidal field. If we relate our effective Love number to this definition, we can identify 
\begin{equation}
    k_2 = \mathrm{Re}(k^\mathrm{eff}_{2})\, ,\quad  \tau_d = \frac{\mathrm{Im}(k_2^\mathrm{eff})}{\mathrm{Re}(k_2^\mathrm{eff})\omega}\, .
\end{equation}
This definition leads to 
 \begin{equation}
     \tau_d \approx -2{\gamma_f \over \sigma_f^2} \, ,
 \end{equation}
which matches the Newtonian definition of the tidal lag in the appropriate limit~\cite{Ghosh_2026,Lai:1993di}. As shown in this work, the imaginary mode frequency $\gamma_f$ changes as a function of temperature when we introduce viscous damping. The damped energy actually converts to thermal energy, changing the overall temperature of the star. Hence, during the inspiral, the parameters $\gamma_f$, as well as $\tau_d$, will not remain constant. To incorporate the temperature dependence, we need to consider the thermal evolution of the star as well, which is not within the scope of this work, but has been addressed in recent works~\cite{Ghosh:2025wfx,Ghosh:2025glz}.\\
\noindent
We also see that $\mathrm{Re}(k^\mathrm{eff}_{l})$ vanishes at the exact mode resonances, and close to the mode resonances, the tidal response is dominated by $\mathrm{Im}(k_2^\mathrm{eff})$. This is why in Fig~\ref{fig:response_g} and ~\ref{fig:response_f}, we see that the impact of viscous dissipation on the tidal response only affects the behaviour of the tidal response close to the mode resonances.

\section{Discussion}
\label{sec:discussion}

In this work, we have developed a framework that consistently incorporates finite chemical equilibration rates into the relativistic stellar perturbation problem. Rather than introducing bulk viscosity through dissipative corrections to the stress-energy tensor, we derive the dissipative response directly from the underlying weak interaction rates and encode it in the perturbation equations through a complex, frequency-dependent dynamical sound speed. This approach provides a unified description of two closely related effects: the composition stratification responsible for buoyancy and the bulk-viscous dissipation arising from finite chemical equilibration time. The formulation naturally converges to the two standard cases: the frozen-composition limit when composition $g-$modes emerge, and the limit of rapid equilibration where buoyancy vanishes. In this work, we focus on the dominant non-leptonic weak interactions involving hyperons, whose equilibration timescales can become comparable to oscillation periods at the low temperatures ($T \sim 10^6 - 10^9$ K) expected in inspiralling neutron stars. By contrast, leptonic weak processes remain significantly slower under these conditions, allowing the associated particle fractions to be treated as effectively frozen. This separation of timescales makes hyperonic reactions the dominant source of chemical equilibration and bulk-viscous dissipation in this relevant temperature regime. \\

Our results demonstrate that finite reaction rates affect the oscillation spectrum in different ways depending on the mode family. For the fundamental $f-$mode, the real oscillation frequency remains essentially unchanged throughout the temperature range, as expected, since the restoring pressure is very weakly affected by the finite reaction rates. The dominant impact instead appears in the imaginary part of the mode frequency. The strongest damping occurs when the mode frequency becomes comparable to the chemical equilibration rate, corresponding to the temperature range where the imaginary part of the dynamical sound speed and the associated bulk viscosity reach their maximum values. For the most massive stars considered here, the viscous damping rate becomes comparable to the gravitational-wave damping rate, reducing the total damping time by up to $30\%$, and viscous dissipation thus leaves a clear imprint on the damping of the modes. This result is particularly interesting because the $f-$mode dominates the dynamical tidal response~\cite{Steinhoff:2016rfi}, and any change in the damping of the $f-$mode by an amount comparable to gravitational-wave losses could therefore leave an imprint on future high-precision gravitational wave observations~\cite{Ghosh:2023vrx,Yu:2025ptm}. \\

The impact on the composition $g-$modes is considerably more dramatic. In the frozen-composition limit, hyperonic stars support relatively high-frequency $g-$modes, with frequencies significantly larger than those typically found in nucleonic matter~\cite{Tran:2022dva}. This originates from the substantial difference between the equilibrium and frozen-composition sound speeds that develops once hyperons start to appear in the stellar core. As the temperature increases, the reaction rates eventually become comparable to the oscillation period itself and the bulk-viscous damping increases by several orders of magnitude. The $g-$modes disappear from the spectrum at temperatures substantially below those required to reach the instantaneous-equilibrium limit. In other words, the modes are removed by viscous damping before the restoring buoyancy force itself vanishes, even before the bulk viscosity reaches its peak. This behavior differs qualitatively from recent studies of chemically equilibrating nucleonic matter~\cite{Zhao:2025pgx} and highlights the much stronger dissipative role played by non-leptonic hyperonic reactions in damping $g-$modes. Note that beyond the bulk viscous peak, the star can have $g-$modes supported by the proton fraction gradient.\\

The stellar oscillations potentially impact the dynamical tidal response of neutron stars before merger. Following the strategy of our recent work~\cite{Andersson_2026}, by matching the interior relativistic perturbations to the near-zone exterior solution, we obtain a frequency-dependent effective Love number that captures both the mode resonances and the dissipative response of the star. The suppression of the $g-$mode resonances with increasing temperature mirrors the disappearance of the modes from the response, while enhanced damping broadens and weakens the $f-$mode resonance. The dissipative part of the response can be interpreted through the \textit{tidal lag}~\cite{Poisson:2009di,Ripley:2023qxo}, which quantifies the phase shift between the external tidal field and the induced quadrupole deformation. In our framework, this lag is directly related to the damping time of the oscillation mode, establishing a direct connection between microscopic weak-interaction processes, macroscopic bulk viscosity, and the dissipative tidal response of neutron stars. Such a connection is particularly appealing because it offers a route through which the composition of dense matter may influence gravitational-wave observables beyond the equilibrium tidal deformability. Although a quantitative assessment of the detectability of tidal lag is beyond the scope of the present work, the results emphasize viscous damping and tidal lag as natural targets for future binary neutron star waveform modeling efforts.\\

Several important extensions remain to be explored both on the stellar perturbation side and in computing the tidal response. The present analysis assumes fixed stellar temperatures. In realistic binaries, the thermal state evolves through cooling and tidal heating, both of which alter the chemical equilibration rates and therefore the strength of bulk-viscous dissipation~\cite{Ghosh:2025glz}. Incorporating thermal evolution self-consistently would allow the tidal lag and mode damping to be followed throughout the inspiral. Another important direction concerns the onset of superfluidity in the stellar core. Firstly, they suppress and modify the weak interaction processes responsible for chemical equilibration~\cite{Haensel_SF_Nu,Haensel_SF_Hyp}.  Secondly, superfluidity introduces additional dynamical degrees of freedom which bring new sets of oscillation modes into existence.  A consistent treatment of superfluid matter within the relativistic perturbation framework is needed to address the impact of viscous dissipation on the superfluid dynamics~\cite{Andersson_2007}. Finally, applying the same framework to other forms of matter, specifically deconfined quark matter~\cite{Ghosh:2025wfx,HegadeKR:2026iou}, would help map out the different composition effects on the dissipation. A particularly promising direction is to map the dissipative tidal response derived within our framework onto the effective dynamical tidal description recently developed using scattering-amplitudes~\cite{Saketh:2026trm}, and incorporating them in a two-body Hamiltonian~\cite{Mandal_2024}. The framework developed here provides a foundation for addressing these questions and for incorporating composition-dependent dissipation effects into future gravitational-wave models of binary neutron-star inspirals.

\section*{Acknowledgements}
The authors thank Debarati Chatterjee, Fabian Gittins, Peter Rau, Yumu Yang and M.V.S. Saketh for valuable suggestions on the manuscript. A.H.~acknowledges financial support by the UKRI under the Horizon Europe Guarantee project EP/Z000939/1. N.A. and S.G. gratefully acknowledge support from the STFC via Grant No. ST/Y00082X/1.

\appendix
\section{Rate Calculation}
\label{App:rate}
In this section, we are deriving the Fermi surface approximation for the strangeness-changing rates used in this paper. This derivation follows earlier calculations, mostly by Refs.~\cite{vanDalen:2003uy,Ofengeim:2019fjy} and Ref.~\cite{Alford2021}.
Using Fermi's golden rule, we can write down the general expression for the rates of all processes:
\begin{widetext}
\begin{align}  
    \Gamma_{12\to 34}= &\frac{1}{S}\int\frac{d^3p_1}{(2\pi)^3}\frac{d^3p_2}{(2\pi)^3}\frac{d^3p_3}{(2\pi)^3}\frac{d^3p_4}{(2\pi)^3}\frac{\sum_{\mathrm{s}}|M_{1234}|^2}{2^4E_1^*E_2^*E_3^*E_4^*}(2\pi)^4\delta\left(E_1+E_2-E_3-E_4\right)\delta^3\left(\mathbf{p_1}+\mathbf{p_2}-\mathbf{p_3}-\mathbf{p_4} \right)\times \label{eq:rateint} \\[2ex]
   &f_1(E_1,\mu_1) f_2(E_2,\mu_2) \left[1-f_3(E_3,\mu_3)\right]\left[1-f_4(E_4,\mu_4)\right] \, , \nonumber
\end{align}  
\end{widetext}
with the symmetry factor $S=2$ for all processes with two identical baryons on one side of the reaction. The spin-summed, squared matrix element of the process $\sum_s|M_{1234}|^2$ is derived in \cite{Alford2021} and given by 
\begin{align}
M_{1234}=&\Large[\bar{u}_3F^S_{23}u_2\,\bar{u}_4F_{14}^Wu_1\,D_{\varphi}(k_1^2) \nonumber \\[2ex]
&-\bar{u_3}F^S_{13}u_1\,\bar{u}_4F_{24}^Wu_2\,D_{\varphi}(k_2^2) \Large] \, . \label{eq:matelem}
\end{align}
Note the normalization of the bispinors following Refs.~\cite{Griffiths:111880} and \cite{Roberts:2016mwj} to $u^{\dagger} u=2E^*$, which leads to the corresponding energy denominators in Eq.~(\ref{eq:rateint}) in terms of
\begin{equation}
    E_i^*=\sqrt{p_i^2+\left(M_i^*\right)^2} \, ,
\end{equation}
with the effective mass $M_i^*=m_i-g_{\sigma i}\langle\sigma\rangle$.
Otherwise, on-shell baryons are characterized by four-momenta that obey the dispersion relation \begin{equation}
    E_i=\sqrt{p_i^2+\left(M_i^*\right)^2}+g_{\omega i}\langle\omega_0\rangle+g_{\rho i}I_{i3}\langle\rho_{03}\rangle +g_{\phi i}\langle\phi\rangle\, .
    \label{eq:fulldisp}
\end{equation}
All meson-field expectation values, such as $\langle\sigma\rangle$, are obtained by solving the Euler-Lagrange equations in the mean-field approximation. For the various coupling constants $g_{\sigma i}$, $g_{\omega i}$, $g_{\rho i}$, and $g_{\phi i}$,  see Ref.~\cite{Gusakov:2014ota}.
The meson propagator $D_\varphi$ depends on the dispersion relations from Eq.~(\ref{eq:fulldisp}), whereas the remaining matrix element is given in terms of $E^*$. The weak and strong interaction vertices are given by
\begin{align}\label{eq:vertex}
    F_{ij}^W= G_F m_\pi^2\left(A_{ij}+B_{ij}\gamma_5 \right)\, , \qquad F_{ij}^S=g_{ij}\gamma_5\, ,
\end{align}
with the Fermi constant $G_F=1.1663787\times 10^{-5}$ GeV$^{-2}$, and the fifth gamma matrix $\gamma_5$. For the strong interaction coupling constants $g_{ij}$ and the weak interaction coupling constants $A_{ij}$ and $B_{ij}$, which depend on the baryons in the corresponding vertex, see Ref.~\cite{Alford2021}. The meson propagator is given by
\begin{equation}
     \label{eq:mesprop}
D_\varphi(k)=\frac{1}{k_0^2-k^2-m_{\varphi}^2} \, .
\end{equation} Depending on the process, the exchanged meson is either a pion (reactions $1-3$ in \ref{eq:reactions}) or a kaon (reactions $4$ and $5$). The phase space factors are described by Fermi-Dirac distribution functions
\begin{equation}
    f_i(E_i,\mu_i)=\frac{1}{1+\exp\left(\frac{E_i-\mu_i}{T}\right)} \, ,
\end{equation}
which are functions of the full dispersion relation of the incoming ($i=1,2$) and outgoing ($i=3,4$) baryons, see Eq.~(\ref{eq:fulldisp}), the chemical potentials $\mu_i$ and the temperature $T$.  
Following Ref.~\cite{Kaminker:2016ayg}, we are splitting the rate integral in Eq.~(\ref{eq:rateint}) into an ``energy'' contribution and an angular contribution, which is justified because of the narrow phase space around the Fermi surface that contributes to the rate at low temperatures. This is known as the ``phase-space decomposition,'' introduced in Ref.~\cite{Shapiro:1983du}. Instead of applying an angular averaging over the matrix element (or even assuming that the matrix element can be described in the ultrarelativistic approximation where it becomes momentum independent), we numerically calculate the angular part and include the angle-dependent matrix element in the angular integral.
The integral can now be written as 
\begin{equation}
    \Gamma=\frac{T^3}{4096\pi^8}\mathcal{I(\xi)}\mathcal{A} \, ,
\end{equation}
where we have defined $\xi=\delta\mu/T$, and the energy integral is given by
\begin{align}
    \mathcal{I}(\xi)&=\prod_{i=1}^4\int dz_i \,k_{F,i}\,f_i\delta(\sum_{j=1}^4 z_j-\xi)\\
    &=k_{F,1}k_{F,2}k_{F,3}k_{F,4}\frac{4\pi^2\xi+\xi^3}{6\left( 1-e^{-\xi}\right)} \, . 
\end{align}
In this equation, we have defined $z_j=(E_j-\mu_j)/T$, and made use of the symmetry properties of the Fermi-Dirac distributions to obtain the desired structure. 
The angular integral is given by
\begin{equation}
    \mathcal{A}=\prod_{i=1}^4\int d\Omega_i \sum_{\mathrm{s}}|M_{1234}|^2 \delta^3(\vec{k}_{F,1}+\vec{k}_{F,2}-\vec{k}_{F,3}-\vec{k}_{F,4})  \, .
\end{equation}
Note that the matrix element is evaluated on the Fermi surface, but still depends on the angles between the participating Fermi momenta. Aligning one of the Fermi momenta with one of the integration axis, and a second one with the plane spanned by the first momentum and the second integration axis, allows us to reduce the initially $8-$dimensional integral to a $5-$d one, out of which three can be performed analytically by integrating over the delta functions. The remaining $2-$d integral is solved numerically. We can calculate the slope of the difference of the back and forward rates $\lambda$ that enters the equilibration time $\gamma$ in Eq.~(\ref{eq:gamma}) from
\begin{equation}
    \Gamma_\leftarrow(\xi)=  \Gamma_\rightarrow(-\xi) \, ,
\end{equation}
which leads to the result stated in Eq.~(\ref{eq:lambda}).
%
\newpage
\renewcommand{\bibsection}{}
\bibliography{ref.bib}

\end{document}